\documentclass[a4paper]{article}

\usepackage{INTERSPEECH2020}

\usepackage{lineno,amsmath,comment,booktabs,hyperref,pgfplots,tikz,epstopdf,multirow,pifont}
\pgfplotsset{compat=1.13}

\newcommand{\cmark}{\ding{51}}%
\newcommand{\xmark}{\ding{55}}%

\makeatletter
\def\bstctlcite#1{\@bsphack
\@for\@citeb:=#1\do{%
\edef\@citeb{\expandafter\@firstofone\@citeb}%
\if@filesw\immediate\write\@auxout{\string\citat
ion{\@citeb}}\fi}%
\@esphack}
\makeatother

\title{Sum-Product Networks for Robust Automatic Speaker Identification}
\name{Aaron~Nicolson~and~Kuldip~K.~Paliwal}
\address{
  Signal Processing Laboratory, Griffith University, Brisbane, Queensland,
Australia, 4111}
\email{aaron.nicolson@griffithuni.edu.au, k.paliwal@griffith.edu.au}

\begin{document}
\bstctlcite{IEEEtrans:BSTcontrol}
\maketitle

\begin{abstract}
We introduce sum-product networks (SPNs) for robust speech processing through a simple robust automatic speaker identification (ASI) task. SPNs are deep probabilistic graphical models capable of answering multiple probabilistic queries. We show that SPNs are able to remain robust by using the marginal probability density function (PDF) of the spectral features that reliably represent speech. Though current SPN toolkits and learning algorithms are in their infancy, we aim to show that SPNs have the potential to become a useful tool for robust speech processing in the future. SPN speaker models are evaluated here on real-world non-stationary and coloured noise sources at multiple signal-to-noise ratio (SNR) levels. In terms of ASI accuracy, we find that SPN speaker models are more robust than two recent convolutional neural network (CNN)-based ASI systems. Additionally, SPN speaker models consist of significantly fewer parameters than their CNN-based counterparts. The results indicate that SPN speaker models could be a robust, parameter-efficient alternative for ASI. Additionally, this work demonstrates that SPNs have potential in related tasks, such as robust automatic speech recognition (ASR) and automatic speaker verification (ASV).
\newline\textbf{Availability}: The SPN ASI system is available at \url{https://github.com/anicolson/SPN-ASI}.	
\end{abstract}

\section{Introduction} \label{seca}
The task of a text-independent automatic speaker identification (ASI) system is to identify a speaker from a given voice recording, irrespective of its linguistic content. This is accomplished by modelling the voice characteristics of each speaker after an enrolment phase \cite{doi:10.1121/1.4964509}. Common applications of ASI include the selection of a speaker-dependent acoustic model for an automatic speech recognition (ASR) system \cite{7918432} and speaker segmentation --- an important pre-processing step for speaker diarisation \cite{Park2018}. The realisation of each application is dependent upon a high-performance ASI system. The first widely adopted ASI system utilised Gaussian mixture model (GMM) speaker models \cite{REYNOLDS199591}.

One obstacle that prevented the commercial introduction of GMM speaker models was their poor performance in the presence of noise \cite{1202292}, spurring the investigation of robust approaches \cite{536825}. A noteworthy approach was the missing-feature approach, which is underpinned by evidence that speech is intelligible to humans even after it has undergone substantial spectral masking \cite{1511828}. \textit{Marginalisation}, as proposed by Cook \textit{et al.} \cite{COOKE2001267}, has been the most prominent missing-feature approach in the literature \cite{5871484}, and is able to significantly increases the robustness of a GMM speaker model \cite{nicolson2018spectral}. For marginalisation, the marginal probability density function (PDF) is obtained by integrating over the components of the feature vector that have been classified as unreliable representations of speech \cite{Nicolson2018}. Classification is thus performed on a partial instantiation of a given feature vector, consisting of only the components that reliably represent speech.

Recently, ASI and automatic speaker verification (ASV) systems employing deep neural networks (DNNs) have demonstrated a higher performance than GMM and i-vector-based systems \cite{6854363}. One example is the x-vector system, which utilises pooling and a DNN trained to discriminate between speakers to map speech to a fixed-size embedding \cite{8461375}. Convolutional neural networks (CNNs) have also been employed \cite{Chung2018}. SincNet is a CNN that employs parametrised sinc functions to pre-define a bank of band-pass filters for its first layer \cite{8639585}. Another example proposed by Xie \textit{et al.} \cite{8683120} utilises a `thin' residual CNN (referred to as Xie2019 henceforth). It also includes dictionary-based NetVLAD \cite{7937898} and GhostVLAD \cite{10.1007/978-3-030-20890-5_3} layers for feature aggregation. Despite their high performance on clean speech, modern ASI systems are still susceptible to performance degradation in the presence of noise \cite{6288857}. Additionally, DNNs are not probabilistic models and cannot employ classifier-compensation missing-feature approaches, such as marginalisation. Currently, the most popular approach to increase the robustness of a DNN-based system is to use a front-end to pre-process the noisy speech \cite{8391748, 10.1145/3178115}.

In 2011, Poon \textit{et al.} \cite{6130310} proposed a deep tractable probabilistic graphical model called the sum-product network (SPN). An SPN can be described as a deep neural network (DNN) restricted to using sum and product operators. When viewed as a probabilistic graphical model, an SPN can be described as a rooted directed acyclic graph with distributions as leaves. SPNs have clear semantics; each node represents an unnormalised joint probability distribution over a set of variables. As they can answer marginal inference queries, SPNs lend themselves well to marginalisation. One disadvantage is that structure and weight learning algorithms for SPNs, as well as libraries, are currently undeveloped, as highlighted by Jaini \textit{et al.} \cite{pmlr-v72-jaini18a}. However, the long-term outlook of SPNs is positive. New SPN toolkits are being developed, such as LibSPN \cite{pronobis2017libspn}, that take advantage of modern machine learning toolkits, such as TensorFlow \cite{tensorflow2015-whitepaper}. Additionally, recently proposed SPN architectures developed for temporal (dynamic SPNs \cite{pmlr-v52-melibari16}) and spatial representations (deep generalised convolutional SPNs (DGC-SPNs) \cite{van2019deep}) have shown promising results.

We propose SPNs and marginalisation for robust ASI. We first formulate marginalisation for SPNs. We then investigate SPNs and marginalisation on a simple robust ASI task. The structure of each SPN speaker model is learned using LearnSPN \cite{pmlr-v28-gens13}. SPN speaker models are evaluated against GMM speaker models \cite{365379}, SincNet \cite{8639585}, and Xie2019 \cite{8683120}. The SPN and GMM speaker models employ marginalisation, whilst SincNet and Xie2019 employ the long short-term memory ideal ratio mask (LSTM-IRM) estimator by Chen \textit{et al.} \cite{doi:10.1121/1.4986931} as a front-end. SPN speaker models are evaluated using multiple conditions, including real-world non-stationary and coloured noise sources and multiple signal-to-noise ratio (SNR) levels. From the presented results, we aim to demonstrate the following: 1) SPN speaker models are more robust than GMM speaker models when marginalisation is and is not used, 2) SPN speaker models utilising marginalisation have the potential to be more robust than recent CNN-based ASI systems that employ a front-end technique, and 3) SPNs and marginalisation have potential in related robust speech processing tasks, such as robust ASR and ASV.



\section{SPN speaker models} \label{sec:a}
\subsection{Features}
For marginalisation, a frequency-domain representation is required. Hence, we employ the log-spectral subband energies (LSSEs) of the clean speech power spectral density (PSD) estimate as features for the SPN and GMM speaker models. The LSSEs are computed from the single-sided PSD estimate:\footnote{For convenience, the time-frame index is omitted from the notation.}
\begin{equation}
\textrm{X}_b = \log{\sum_{k=0}^{N_d/2} h_{b,k} \hat P_k, \quad 0\leq b \leq B-1}, 
\end{equation}
where $N_d$ denotes the time-frame duration in discrete-time samples, $k$ denotes the discrete-frequency bin, $\hat P_k$, for all $k$, denotes the PSD estimate for a given time-frame, and $h_{b,k}$, for all $k$, denotes the $b^{th}$ filter of a bank of $B$ triangular-shaped critical band filters spaced uniformly on the mel-scale. The PSD is estimated from the short-time Fourier transform (STFT) of the clean speech using the periodogram method, as in \cite{nicolson2018spectral}.


\begin{figure} [ht]
	\begin{center}
		\includegraphics[scale=0.45]{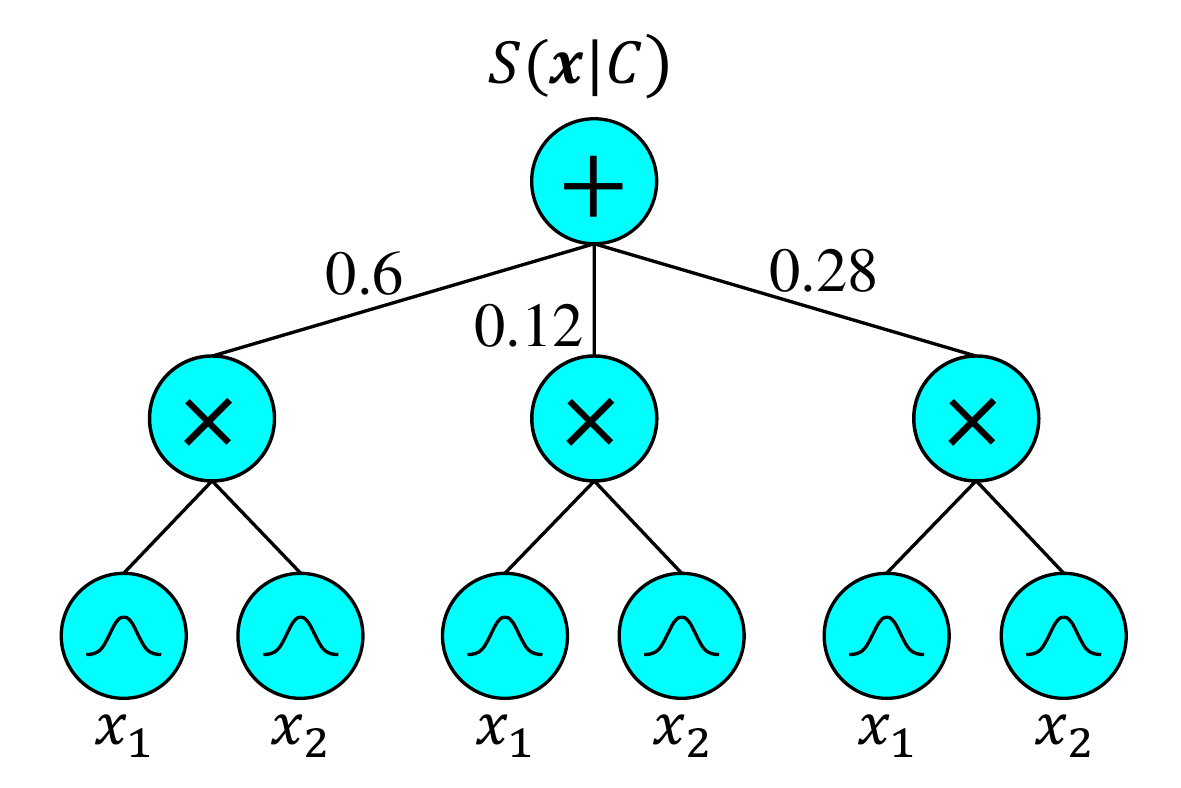}
		\caption{SPN speaker model with univariate Gaussian leaves.}
		\label{fig:SPN1} 
	\end{center}
\end{figure}

\subsection{SPN speaker models with Gaussian leaves}
An SPN \cite{6130310} specifies an unnormalised joint distribution over a set of random variables,  $\textbf{X}=(X_1,X_2,...,X_B)^\top$, where in this case, $\textbf{X}$ is the LSSEs for a time-frame of clean speech. An observation of $\textbf{X}$ is denoted by $\textbf{x}=(x_1,x_2,...,x_B)^\top$. Hence, the SPN, $S$, for speaker class $C$ is a function of the observed feature vector, $S(\textbf{x}|C)$, where the value of the SPN is given by its root. An SPN consists of multiple layers of sum and product nodes, with distributions as leaves. The multivariate distribution of the $i^{th}$ leaf is over a subset of the variables: $\textbf{X}_i \subseteq \textbf{X}$, and is assumed to be normally distributed: $\textbf{X}_i \sim \mathcal{N}(\boldsymbol{\mu},\boldsymbol{\Sigma}|i,C)$, with mean $\boldsymbol{\mu}$, and diagonal covariance $\boldsymbol{\Sigma}$. The PDF for the $i^{th}$ leaf is given by

\begin{equation}
\mathcal{N}(\textbf{x}_i|i,C)=\prod_{d\in \textbf{D}} \frac{1}{\sqrt{2\pi{\Sigma_{i,C}(d,d)} }}~e^{-\frac{(\textbf{x}_i(d) - {\boldsymbol \mu_{i,C}(d)})^2}{2\Sigma_{i,C}(d,d)}}, 
\end{equation}
where $\textbf{D} \subseteq (1,2,...,B)^\top$ indicates the random variable indices for $\textbf{X}_i$. An SPN over two variables with univariate Gaussian leaves is shown in Figure \ref{fig:SPN1}. 

If node $i$ is a product node, its value is given by the product of the values of its children, $Ch(.)$: $S_i = \prod\nolimits_{j \in Ch(S_i)}S_j$, where $S_j$ is the $j^{th}$ child of node $S_i$. If node $i$ is a sum node, its value is given by the sum of the values of its children: $S_i = \sum\nolimits_{j \in Ch(S_i)} w_{ij}S_j$, where weight $w_{ij}$ is the non-negative weighted edge between $S_i$ and $S_j$. To be a \textit{valid} joint distribution, an SPN must be both \textit{decomposable}, and \textit{complete}, as described in \cite{6130310}. The scope of a node, $Sc(.)$, is defined as the set of variables that are descendants of it. An SPN is said to be decomposable when the scopes of the children of its product nodes are disjoint: $\forall S_j, S_k \in Ch(S_i), Sc(S_j) \cap Sc(S_k)=\emptyset$, where $\emptyset$ indicates an empty set. An SPN is said to be complete when the scopes of the children of its sum nodes are identical: $\forall S_j, S_k \in Ch(S_i),~Sc(S_j)=Sc(S_k)$.


\begin{table*}[ht]
	\centering
    \setlength{\tabcolsep}{3.6pt}
	\caption{ASI accuracy ($\%$) for the real-world non-stationary noise sources. The average improvement over the model in the preceding row is shown in the last column. The highest accuracy for each condition is shown in boldface.}
	
	\begin{tabular}{lll|lllll|lllll|l} 
		\toprule
		\multirow{3}{*}{\begin{tabular}[c]{@{}c@{}}\textbf{Model}\end{tabular}} & \multirow{3}{*}{\begin{tabular}[c]{@{}c@{}}\textbf{Marg.}\end{tabular}} & \multirow{3}{*}{\begin{tabular}[c]{@{}c@{}}\textbf{Bounds}\end{tabular}} & 
		\multicolumn{10}{c|}{\bf SNR level (dB)} & 
		\multirow{3}{*}{\begin{tabular}[c]{@{}c@{}}\textbf{Average}\\ \textbf{impr.}\end{tabular}} \\ 
		\cline{4-13}
		& & & \multicolumn{5}{c|}{\bf Voice babble} & \multicolumn{5}{c|}{\bf Street music} \\ 
		\cline{4-13}
		& & & {\bf-5}  & {\bf0}   & {\bf5}   & {\bf10}  & {\bf15}        & {\bf-5}  & {\bf0}   & {\bf5}   & {\bf10}  & {\bf15} \\ 
		\hline
		
		GMM \cite{365379} & \xmark & \xmark & \textbf{0.00} & \textbf{0.00} & 0.63 & 13.02 & \textbf{50.48} & 0.00 & \textbf{0.00} & 0.95 & 5.40 & 25.40 & - \\
		SPN & \xmark & \xmark & \textbf{0.00} & \textbf{0.00} & \textbf{1.59} & \textbf{15.56} & 50.16 & \textbf{0.00} & \textbf{0.32} & \textbf{1.27} & \textbf{6.03} & \textbf{25.71} & +0.48 \\\midrule 
		GMM \cite{365379} & \cmark & \xmark & \textbf{2.22} & 6.35 & 18.10 & 46.98 & 79.37 & \textbf{4.76} & \textbf{10.48} & 20.32 & 37.46 & 66.35 & - \\
		SPN & \cmark & \xmark & \textbf{2.22} & \textbf{7.30} & \textbf{19.05} & \textbf{50.79} & \textbf{83.49} & 4.13 & \textbf{10.48} & \textbf{24.13} & \textbf{40.95} & \textbf{71.43} & +2.16 \\\midrule 
		GMM \cite{365379} & \cmark & \cmark & \textbf{15.24} & 29.21 & 48.57 & 72.70 & 89.21 & 20.63 & 32.06 & 54.60 & 71.11 & 85.40 & - \\
		SPN & \cmark & \cmark & 14.60 & \textbf{32.70} & \textbf{55.87} & \textbf{77.78} & 91.43 & \textbf{22.54} & \textbf{34.92} & \textbf{59.37} & \textbf{74.29} & 90.16 & +3.49 \\
		SincNet \cite{8639585} + IRM \cite{doi:10.1121/1.4986931} & - & - & 0.63 & 4.44 & 25.40 & 71.75 & 92.70 & 1.27 & 5.40 & 23.81 & 64.44 & \textbf{92.38} & -17.14 \\
		SincNet \cite{8639585} & - & - & 0.32 & 1.59 & 18.10 & 56.83 & \textbf{93.02} & 0.63 & 2.86 & 11.11 & 46.98 & 85.40 & -6.54 \\
		Xie2019 \cite{8683120} + IRM \cite{doi:10.1121/1.4986931} & - & - & 0.63 & 1.27 & 10.48 & 28.89 & 53.33 & 0.32 & 1.27 & 4.44 & 20.63 & 40.95 & -15.46 \\
		Xie2019 \cite{8683120} & - & - & 0.32 & 0.95 & 4.13 & 14.92 & 41.27 & 0.00 & 0.32 & 2.54 & 13.65 & 35.56 & -4.85 \\
		
		\bottomrule          
	\end{tabular}
	\label{tab:1}
\end{table*}

\subsection{Marginalisation for SPNs}
For marginalisation, each component of an observed noisy speech feature vector is classified as either a reliable or an unreliable representation of the corresponding unobserved clean speech component. The noisy speech feature vector, $\textbf{y}$, can thus be described as the union of the reliable and unreliable components: $\textbf{y}= \textbf{y}^r \cup \textbf{y}^u$. Here, we not only apply marginalisation to SPNs, but also \textit{bounded marginalisation}, as proposed by Cook \textit{et al.} \cite{6130310}. For bounded marginalisation, the value of an unreliable component is utilised as the upper bound of the unobserved clean speech component value. For LSSEs, the bounds are taken from $[-\infty,\textbf{y}_n^u]$. Thus, the PDF for the $i^{th}$ leaf becomes:
\begin{equation}
\label{equ:R1}
\mathcal{N}(\textbf{y}_i^r, \textbf{x}_i^u \leq \textbf{y}_i^u|i,C)=\mathcal{N}(\textbf{y}_i^r|i,C) \int_{- \infty}^{\textbf{y}_i^u} \mathcal{N}(\textbf{x}_i^u |i,C) d\textbf{x}_i^u.
\end{equation}
For marginalisation, the unreliable components are treated as missing and the bounds are taken from $[-\infty,\infty]$. The integral in Equation (\ref{equ:R1}) thus reduces to unity, giving $\mathcal{N}(\textbf{y}_i^r|i,C)$. When all of the components of ${\bf y}_i$ are unreliable, it is treated as a vector with no instantiated components: $\mathcal{N}(\textbf{y}_i^r= \emptyset|i,C)=1$. 

\section{Experiment setup} \label{sec:b}

\subsection{Signal processing}
The feature vectors for the GMM and SPN speaker models are computed using a Hamming window function, with a time-frame duration of 32 ms (512 discrete-time samples) and a time-frame shift of 16 ms (256 discrete-time samples). The 257-point single-sided PSD estimate for a time-frame is used and includes both the DC and Nyquist frequency component. The LSSEs are computed from the PSD estimate using 26 triangular-shaped critical band filters spaced uniformly on the mel-scale.

\subsection{Classification of reliable spectral components}
Here, the reliability of a spectral component is determined by its \textit{a priori} SNR, as in \cite{Barker2000}. A component with an \textit{a priori} SNR greater than 0 dB is classified as reliable \cite{Wang2005}. Deep Xi-ResNet from \cite{9066933} is used here as the \textit{a priori} SNR estimator. Deep Xi is a deep learning approach to \textit{a priori} SNR estimation \cite{NICOLSON201944}, and is available at: \url{https://github.com/anicolson/DeepXi}. It estimates the \textit{a priori} SNR for each of the 257 frequency-domain components of a noisy speech time-frame. The \textit{a priori} SNR estimate for each subband is subsequently found by applying the filterbank used to compute the LSSEs.

\subsection{Training and test sets}
The TIMIT corpus \cite{garofolo1993darpa} ($16$ kHz, single-channel), which consists of $630$ speakers with $10$ utterances each, is used as the clean speech. The $si^*$ and $sx^*$ subsets are used for training ($5\,040$ utterances) and the $sa^*$ subset is used for testing ($1\,260$ utterances). Each clean speech recording from the $sa^*$ subset is mixed additively with one of four real-world noise source recordings to create the noisy speech for testing ($315$ clean speech recordings for each noise source). Each noisy speech recording is replicated at five SNR levels: $\{-5,0,5,10,15\}$ dB, forming a test set of $6\,300$ noisy speech recordings. The real-world noise sources include two non-stationary and two coloured. The two real-world non-stationary noise sources include \textit{voice babble} from the RSG-10 noise dataset \cite{steeneken1988description} and \textit{street music} (recording no. $26\,270$) from the Urban Sound dataset \cite{10.1145/2647868.2655045}. The two real-world coloured noise sources include \textit{F16} and \textit{factory} (welding) from the RSG-10 noise dataset \cite{steeneken1988description}. 

\subsection{ASI systems}

\hspace{\parindent} \textbf{GMM:} For each speaker, a GMM consisting of $48$ diagonal-covariance clusters is trained on the training set using the expectation-maximisation (EM) algorithm \cite{doi:10.1111/j.2517-6161.1977.tb01600.x}, and the k-means++ algorithm for parameter initialisation \cite{ilprints778}. 

\textbf{SincNet:} \cite{8639585} is available at: \url{https://github.com/mravanelli/SincNet} and is trained using the training set with default hyperparameters. 
	 
\textbf{Xie2019:} \cite{8683120} is available at: \url{https://github.com/WeidiXie/VGG-Speaker-Recognition} and is trained using the training set with default hyperparameters and a 1 second input spectrogram size.	 
	 
\textbf{SincNet + IRM \& Xie2019 + IRM:} The LSTM-IRM estimator from \cite{doi:10.1121/1.4986931} is used as the front-end for SincNet and Xie2019. The training data and configuration from \cite{nicolson2019deep} is used specifically.
	
\textbf{SPN:} Each speaker is modelled using an SPN with univariate Gaussian leaves. The SPFlow library is used to implement the SPN speaker models \cite{molina2019spflow}. A variant of the LearnSPN algorithm \cite{pmlr-v28-gens13} that partitions and clusters variables using the Hirschfeld-Gebelein-R\'enyi maximum correlation coefficient \cite{AAAI1816865} is used as the structure learning algorithm. The minimum number of instances to split is set to $50$ and the threshold of significance is set to $0.3$ for the structure learning algorithm.
	
\begin{table*}[ht]
	\centering
	\setlength{\tabcolsep}{3.6pt}
	\caption{ASI accuracy ($\%$) for the real-world coloured noise sources. The average improvement over the model in the preceding row is shown in the last column. The highest accuracy for each condition is shown in boldface.}
	
	\begin{tabular}{lll|lllll|lllll|l} 
		\toprule
		\multirow{3}{*}{\begin{tabular}[c]{@{}c@{}}\textbf{Model}\end{tabular}} & \multirow{3}{*}{\begin{tabular}[c]{@{}c@{}}\textbf{Marg.}\end{tabular}} & \multirow{3}{*}{\begin{tabular}[c]{@{}c@{}}\textbf{Bounds}\end{tabular}} & 
		\multicolumn{10}{c|}{\bf SNR level (dB)} & 
		\multirow{3}{*}{\begin{tabular}[c]{@{}c@{}}\textbf{Average}\\ \textbf{impr.}\end{tabular}} \\ 
		\cline{4-13}
		& & & \multicolumn{5}{c|}{\bf F16} & \multicolumn{5}{c|}{\bf Factory} \\ 
		\cline{4-13}
		& & & {\bf-5}  & {\bf0}   & {\bf5}   & {\bf10}  & {\bf15}        & {\bf-5}  & {\bf0}   & {\bf5}   & {\bf10}  & {\bf15} \\ 
		\hline
		
		GMM \cite{365379} & \xmark & \xmark & \textbf{0.32} & \textbf{0.32} & \textbf{0.95} & 0.95 & 10.16 & \textbf{0.63} & \textbf{1.27} & \textbf{0.63} & 1.90 & 12.06 & - \\
		SPN & \xmark & \xmark & \textbf{0.32} & \textbf{0.32} & 0.32 & \textbf{2.54} & \textbf{14.92} & \textbf{0.63} & 0.63 & \textbf{0.63} & \textbf{2.54} & \textbf{13.65} & +0.73 \\\midrule 
		GMM \cite{365379} & \cmark & \xmark & \textbf{1.90} & 7.30 & 21.27 & 34.29 & 58.73 & \textbf{3.17} & 5.71 & 10.79 & 25.40 & 53.65 & - \\
		SPN & \cmark & \xmark & \textbf{1.90} & \textbf{10.16} & \textbf{21.59} & \textbf{34.60} & \textbf{59.37} & 2.54 & \textbf{6.35} & \textbf{14.29} & \textbf{28.89} & \textbf{55.87} & +1.33 \\\midrule 
		GMM \cite{365379} & \cmark & \cmark & 19.37 & 35.24 & 46.98 & 62.54 & 80.32 & \textbf{11.75} & 18.41 & 36.83 & 54.60 & 81.90 & - \\
		SPN & \cmark & \cmark & \textbf{22.54} & \textbf{36.83} & \textbf{49.84} & \textbf{66.35} & \textbf{81.90} & 10.48 & \textbf{21.59} & \textbf{39.68} & \textbf{56.83} & 82.54 & +2.06 \\
		SincNet \cite{8639585} + IRM \cite{doi:10.1121/1.4986931} & - & - & 0.63 & 1.27 & 5.71 & 26.67 & 72.70 & 0.95 & 1.59 & 13.02 & 44.13 & \textbf{86.67} & -21.52 \\
		SincNet \cite{8639585} & - & - & 0.32 & 0.63 & 4.13 & 16.19 & 57.78 & 0.00 & 0.95 & 5.71 & 35.56 & 78.41 & -5.37 \\
		Xie2019 \cite{8683120} + IRM \cite{doi:10.1121/1.4986931} & - & - & 0.32 & 0.32 & 2.86 & 6.98 & 20.00 & 0.00 & 0.32 & 0.63 & 2.86 & 21.27 & -14.41 \\
		Xie2019 \cite{8683120} & - & - & 0.32 & 0.63 & 3.17 & 7.62 & 21.90 & 0.95 & 0.63 & 1.27 & 5.71 & 26.67 & +1.33 \\
		
		\bottomrule          
	\end{tabular}
	\label{tab:2}
\end{table*}
	
\section{Results and discussion} \label{sec:c}

\subsection{Real-world non-stationary noise sources}

Table \ref{tab:1} shows the ASI accuracy for two real-world non-stationary noise sources: \textit{voice babble} and \textit{street music}. Over all of the tested conditions in Table \ref{tab:1}, SPN speaker models demonstrated an average improvement of $0.48\%$ over GMM speaker models (no marginalisation). This indicates that SPN speaker models are better able to model the joint distribution of each speaker's features. It can be seen that the robustness of SPN speaker models increases significantly when either marginalisation or bounded marginalisation is used. SPN speaker models attained an average improvement of $2.16\%$ and $3.49\%$ over GMM speaker models when marginalisation and bounded marginalisation are used, respectively. The performance improvement that SPN speaker models posses over GMM speaker models is thus extended when either marginalisation or bounded marginalisation is used.

SPN speaker models employing bounded marginalisation are able to outperform SincNet + IRM, with an average improvement of $17.14\%$. While SincNet + IRM achieved the best accuracy at 15 dB for both non-stationary noise sources, it is outperformed at lower SNR levels by SPN speaker models employing bounded marginalisation. The results presented in Table \ref{tab:1} show that SPN speaker models are robust to real-world non-stationary noise sources when marginalisation and bounded marginalisation is used, especially at lower SNR levels. 


\begin{table}[ht]
	\centering
	\setlength{\tabcolsep}{3.6pt}
	\caption{Average number of parameters used by each ASI system for each of the 630 speakers.}
	\begin{tabular}{@{}lllll@{}}
		\toprule
		& \textbf{SPN} & \textbf{GMM} & \textbf{Xie2019} & \textbf{SincNet} \\ \midrule
		\textbf{Params. per speaker} & $2\,502$         & $2\,544$         & $13\,545$            & $36\,718$            \\ \bottomrule
	\end{tabular}
	\label{tab:3}
\end{table}

\subsection{Real-world coloured noise sources}

Table \ref{tab:2} shows the ASI accuracy for two real-world coloured noise sources: \textit{F16} and \textit{factory}. Over all of the tested conditions, SPN speaker models demonstrated an average improvement of $1.33\%$ and $2.66\%$ over GMM speaker models when marginalisation and bounded marginalisation are used, respectively. This indicates that marginalisation and bounded marginalisation are more suited to SPN speaker models than GMM speaker models. SPN speaker models utilising bounded marginalisation were also able to outperform SincNet + IRM, with an average performance increase of $21.52\%$.

The results presented in Tables \ref{tab:1} and \ref{tab:2} show that SPN speaker models are robust to both real-world non-stationary and coloured noise sources when marginalisation or bounded marginalisation is used. The number of parameters that each ASI system expends on a speaker is specified in Table \ref{tab:3}. SPN speaker models are more robust than SincNet, whilst employing 14.7 times fewer parameters on average per speaker. This exhibits the parameter efficiency of SPN speaker models.

\subsection{Future direction}

In this work, standard SPNs are used as speaker models. The structure learning algorithm used to find each SPN speaker model is LearnSPN (introduced in 2013) \cite{pmlr-v28-gens13}, which was the second-ever proposed. An increase in performance can likely be realised by utilising more advanced SPN architectures, such as random and tensorised SPNs (RAT-SPNs) \cite{peharz2018probabilistic}, DGC-SPNs \cite{van2019deep}, or dynamic SPNs \cite{pmlr-v52-melibari16}. Such SPN architectures use predefined structures (i.e. no structure learning algorithm is required) and can be trained discriminatively using modern stochastic gradient descent optimisation algorithms \cite{ruder2016overview}. Alternatively, they can be trained generatively using hard EM \cite{6130310}. These capabilities are to be made available to researchers through toolkits such as LibSPN \cite{pronobis2017libspn}. In this work we investigate SPNs and marginalisation on a simple robust ASI task. ASI was chosen as the task to demonstrate the robustness capabilities of SPNs, as a system could be quickly developed. The results presented in this work lead to more complicated robust speech processing tasks being investigated in future work (complicated in the sense that more system development is required). Such tasks include robust ASR and ASV. 


\section{Conclusion} \label{sec:d}

SPNs utilising marginalisation are proposed for robust ASI. They are evaluated using real-world non-stationary and coloured noise sources at multiple SNR levels. It was found that SPN speaker models and marginalisation are more robust than two recent CNN-based ASI systems that employ significantly more parameters. With the development of new toolkits and architectures, SPNs and marginalisation are predicted to have a bright future in robust ASI, as well as robust ASR and ASV.

\bibliographystyle{IEEEtran}
\bibliography{ms}

\end{document}